\documentclass[a4paper,11pt]{article}
\usepackage{pos}

\title{Remarks on brane-antibrane inflation}
\author*[a]{Gonzalo Villa}

\affiliation[a]{DAMTP,  Centre for Mathematical Sciences,  University of Cambridge, \\ Wilberforce Road,  Cambridge, CB3 0WA, UK}

\emailAdd{gv297@cam.ac.uk}

\abstract{We provide an accessible review to the eta problem in brane-antibrane inflation and a recently proposed solution.
The description of the antibrane by means of nonlinearly realized supersymmetry suggests a simple stabilization mechanism for the volume mode, based on the most general Kahler and superpotentials compatible with the symmetries of the problem.
This work is a contribution to the proceedings of the second general meeting of the COST Action CA21106 (Cosmic WISPers) and is based on~\cite{Cicoli:2024bwq}.}

\FullConference{2nd Training School and General Meeting of the COST Action COSMIC WISPers  (CA21106) (COSMICWISPers2024)\\
 10-14 June 2024 and 3-6 September 2024\\
Ljubljana (Slovenia) and Istanbul (Turkey)\\}

\begin{document}
\maketitle

\section{Inflation and the UV}
Cosmic inflation remains the standard realisation of the early Universe.
In addition to its well known attractive features in addressing the flatness and horizon problems and explaining the origin of cosmic structure, a particularly interesting aspect of inflation is its sensitivity to high-energy physics.
The reason is that simple realizations of the scenario require very flat scalar potentials, along which a scalar field slowly rolls, yielding a quasi-de Sitter background.
Such shapes for the scalar potentials are, of course, sensitive to corrections.

The problem can be divided into two parts (following~\cite{Baumann:2014nda}): in the renormalizable Lagrangian, the mass of the inflaton must be protected by a symmetry in order not to be pushed to the cutoff of the theory.
Typical examples of such symmetries in string model building are shift, scaling, or supersymmetry (SUSY)~\cite{Burgess:2020qsc}, and the three of them play a role in brane-antibrane inflation.
More complicated is the non-renormalizable part.
The flatness of the scalar potential is sensitive to the existence of high dimension operators, inherited from the UV.
Consider for instance the following correction to a given scalar potential $V(\phi)$:
\begin{equation}\label{eq:potential}
V(\phi)=V_0(\phi)\left( 1+ \left( \frac{\phi}{\Lambda}\right)^2 \right) \, .
\end{equation}
Here we have assumed an expansion in powers of $\phi/\Lambda$.
Even though the correction to the overall scale of the potential is small, there is a large contribution to the second slow-roll parameter:
\begin{equation}\label{eq:eta-sr}
\eta=M_p^2\left(\frac{V''(\phi)}{V(\phi)}\right) \to \left(\frac{M_p}{\Lambda}\right)^2\, ,
\end{equation}
where in the last step we have used the potential~\eqref{eq:potential} and assumed that the contribution from $V_0(\phi)$ to $\eta$ is small.
This correction is therefore dangerous generically.

It is thus important to embed models of cosmic inflation into UV complete models.
In this regard, brane-antibrane inflation stands out as particularly interesting.
On top of being favoured by observations, its scalar potential can be tuned exponentially flat by choice of microscopic parameters of order one, and its inherent stringyness carries over to the reheating period, where strong departures from effective field theory are expected.
We will not attempt to justify these facts in this summary article, which is instead intended to address the issues of these models.
These include the spontaneous breaking of supersymmetry by the antibrane, the need to describe a mass-acquisition mechanism for the volume modulus (the WISP by excellence in string compactifications) and the eta problem.
We henceforth work in Planck units $M_p=1$.

\section{Brane-antibrane inflation and the eta problem}

Let us describe the brane-antibrane potential~\cite{Dvali:1998pa,Burgess:2001fx}.
In low-energy string vacua, brane positions are described by scalar fields.
Branes and antibranes are dubbed as such because they carry opposite charges under certain gauge forces whose details are not relevant for our current purposes.
What is important is that these charges induce forces among the objects, described by a Coulomb potential:
\begin{equation}\label{eq:coulomb}
V(\phi , \mathcal{V})=C_0 \left( 1-\frac{D_0}{\phi^4}\right)\, ,
\end{equation}
where $\phi$ is the canonically normalized inter-brane separation, which acts as the inflaton.
When the antibrane sits in a so-called \textit{warped throat}, the functions $C_0$ and $D_0$ are of the form
\begin{equation}
C_0 \sim D_0 \sim \frac{e^{-8\pi K/3g_s M}}{\mathcal{V}^{4/3}} \, , \qquad K\, , \, M \, \in \mathbb{N}\, , g_s \ll 1\, .
\end{equation}
The potential in Eq.~\eqref{eq:coulomb} would be ideally suited for inflation if it was not for its dependence on the field $\mathcal{V}$, the \textit{volume} of the compactification.

\subsection{The volume modulus and the eta problem}
The volume modulus is a generic feature of higher-dimensional theories: degeneracies in their space of solutions manifest themselves at low energies as four-dimensional scalar fields with no tree-level potential, generating a scaling symmetry.
In the case of interest, the antibrane induces a steep scalar potential for the volume modulus, spoiling inflation.
Thus, the potential in Eq.~\eqref{eq:coulomb} only drives inflation if the expectation value of this modulus is fixed (\textit{moduli stabilization}).

The stabilization of the volume mode was first addressed in~\cite{Kachru:2003sx} and leads to the incarnation of the eta problem in this model.
To see this, we note that the physical volume $\mathcal{V}$ is given by the sum of a geometric contribution $\rho$ and a $\phi$-dependent correction due to the backreaction of the brane in the internal space:
\begin{equation}
\mathcal{V}^{2/3}=\rho \left(1 -\bar{\phi}\phi \right)\, .
\end{equation}
This structure can be understood in terms of a generalized shift symmetry~\cite{Burgess:2020qsc}.
In the original proposal, only the geometric contribution $\rho$ is stabilized.
This generates an eta problem, since
\begin{equation}
V(\mathcal{V},\phi) \sim \frac{1}{\left( \langle \rho \rangle(1 -\bar{\phi}\phi )\right)^2}\left( 1-\frac{D_0}{\phi^4}\right) \sim \frac{1}{\langle \rho \rangle^2}\left( 1-\frac{D_0}{\phi^4}\right) \left(1+2 \bar{\phi}\phi \right)\, .
\end{equation}
This potential is of the form of Eq.~\eqref{eq:eta-sr} and generates an order one correction to $\eta$.
The discussion, however, also illustrates how to avoid the eta problem: it is $\mathcal{V}^{2/3}$ that needs to be stabilized\footnote{For aficionados: the approach in~\cite{Kachru:2003sx} used non-perturbative corrections to the superpotential (which by holomorphy cannot depend on $\bar{\phi}\phi$) to stabilize $\rho$. The way out is to consider perturbative corrections to the Kahler potential. }.

\subsection{Nonlinearly realized SUSY to the rescue}
Another issue to consider is the usage of the effective action in the presence of the antibrane.
This object breaks supersymmetry, potentially generating corrections that could spoil inflation (and presumably the mechanism that stabilises the volume modulus).
A way to take into account the effect of these corrections in a systematic manner whilst preserving the powerful properties of supersymmetry (such as holomorphy of the superpotential) has been developed in recent years, in the language of nonlinearly realized supersymmetry.
We will not dwell in the details in this article, simply noting that the antibrane at the bottom of the warped throat can be described in this formalism~\cite{Kallosh:2015nia}.
In a nutshell, the perturbative scalar potential is given by five ingredients:
\begin{itemize}
\item Three functions of the volume\footnote{The generalized shift symmetry that allows these functions to depend only in the combination $\mathcal{V}^{2/3}=\rho-\bar{\phi}\phi$ is expected to hold to all orders in $\alpha'$, but string loop corrections render additional dependence on $\phi$. In this work we assume that the leading dependence of $g(x)$ in $\rho$ is so that the symmetry is respected. Additional $\phi$ dependence is negligible.} $f(\mathcal{V}^{2/3})$, $g(\mathcal{V}^{2/3})$ and $h(\mathcal{V}^{2/3})$ which at tree level satisfy $f(x)=x$, $g(x)=0$ and $h(x)=1$ and admit an expansion in powers (and possibly logarithms) of $1/x$.
These form the Kahler potential.

\item A (for our current purposes) tunable constant $W_0$ and a function of $\phi$ but not of $\mathcal{V}$ which contains the Coulomb potential, $W_x$.
These form the superpotential.
\end{itemize}

The most general perturbative scalar potential built out of these functions is of the form~\cite{Cicoli:2024bwq}:
\begin{equation}\label{eq:potential-general}
V=\frac{1}{U}\left[\left( f' W_x-3g'W_0\right)^2-f'' \left(f W_x^2-6g\, W_0 W_x-9h W_0^2 \right) \right] \, ,
\end{equation}
where $U$ is a function of $f$, $g$ and $h$.
The tree-level contribution reproduces Eq.~\eqref{eq:coulomb}, but the inclusion of corrections reveals an interesting stabilization mechanism.
Assuming a hierarchy between the two terms and neglecting the second, the potential is a perfect square which features a minimum at
\begin{equation}\label{eq:v-stable}
\langle \mathcal{V} \rangle \sim \left( \frac{W_0}{W_x}\right)^{1/a}\, ,
\end{equation}
where we have assumed that $g'(x)\sim x^{-a}$.
Remembering that $W_x$ is naturally exponentially small, this contribution fixes the vacuum expectation value of the volume modulus at exponentially large (in $K/g_s M$) values.
The remaining contribution to the potential in Eq.~\eqref{eq:potential-general} then provides a positive\footnote{Note $f''(x)$ is proportional to the well-known BBHL correction, whose sign depends in the topology of the manifold.} contribution which leads to a potential of the form~\eqref{eq:coulomb}, as discussed in~\cite{Cicoli:2024bwq}.

\section{An example and concluding remarks}

\begin{figure}[t!]
\centering
    \includegraphics[width=0.4\textwidth]{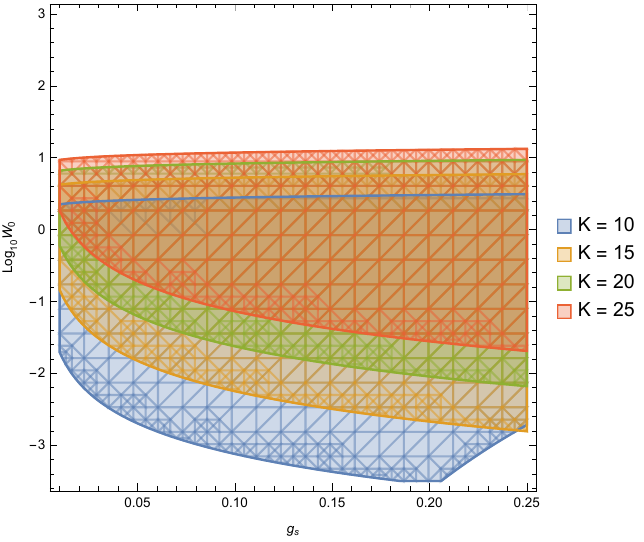}
    \caption{Microscopic regions of parameter space compatible with theoretical and experimental constraints for $g(\sigma)=g_s \log (\mathcal{V}^{2/3})$.}
    \label{fig:plot}

\end{figure}

We have seen that the framework of nonlinearly realized supersymmetry reveals a perfect square structure capable of stabilizing the volume modulus.
In Fig.~\ref{fig:plot} we illustrate\footnote{The plots also take into account experimental and theoretical constraints, such as control over the effective description.} the allowed parameter space for an example function $g(x) \sim \log (x)$, inspired by a 1-loop logarithmic correction.
It is worth noting, however, that Eq.~\eqref{eq:v-stable} is more general and only assumes a hierarchy of the two terms in Eq.~\eqref{eq:potential-general}.
In absence of such a hierarchy, the scalar potential features more ingredients and stabilization is expected to occur - we presented another possibility for $g(x)=0$ in~\cite{Cicoli:2024bwq}.
The existence of a correction which provides a nonzero value for $g(x)$ - as expected on general grounds of effective field theory, since no symmetry argument forbids its presence - deserves further study.

In addition to proposing two different solutions to the eta problem, we also considered post-inflationary dynamics in~\cite{Cicoli:2024bwq}: after brane-antibrane annihilation the scalar potential is modified, and the volume modulus rolls towards a late-time minimum, affecting fundamental constants as it varies, in line with~\cite{Apers:2024ffe}.
Furthermore, inflation ends via brane-antibrane annihilation, releasing their energy into highly excited fundamental strings.
Depending on their decay channels, these might form a network of cosmic superstrings~\cite{Sarangi:2002yt} or enter a Hagedorn phase~\cite{Frey:2024jqy} - in both cases the phenomenology (in particular as gravitational waves) is rich and stringy.
Brane-antibrane inflation thus remains as an attractive candidate for inflation with interesting phenomenology.

\section*{Acknowledgements}

I thank Michele Cicoli, Chris Hughes, Ahmed Kamal, Francesco Marino, Fernando Quevedo and Mario Ramos Hamud for their collaboration in this project and Anshuman Maharan for helpful discussions.
This work has been partially supported by STFC consolidated grant ST/T000694/1.
This article is based upon work from COST Action COSMIC WISPers CA21106, supported by COST (European Cooperation in Science and Technology).

\bibliographystyle{utphys}
\bibliography{biblio}

\end{document}